\documentclass[aps,prx,reprint,superscriptaddress,amsmath,amssymb]{revtex4-2}

\usepackage[T1]{fontenc}
\usepackage[utf8]{inputenc}
\usepackage{lmodern}
\usepackage[english]{babel}

\usepackage[colorlinks=true]{hyperref}
\usepackage[capitalize,nameinlink]{cleveref}
\usepackage{natbib}
\usepackage{graphicx}
\graphicspath{{figs/}}
\usepackage{braket}
\usepackage{siunitx}

\begin{document}


\title{Transconductance quantization in a topological Josephson tunnel junction circuit}


\author{L. Peyruchat}
\affiliation{$\Phi_{0}$, JEIP, USR 3573 CNRS, Collège de France, PSL University, 11, place Marcelin Berthelot, 75231 Paris Cedex 05, France}
\author{J. Griesmar}
\altaffiliation[Current address: ]{Institut Quantique et Département GEGI, Université de Sherbrooke, Sherbrooke, QC, Canada}
\author{J.-D. Pillet}
\altaffiliation[Current address: ]{LSI, CEA/DRF/IRAMIS, Ecole Polytechnique, CNRS, Institut Polytechnique de Paris, F-91128 Palaiseau, France}
\author{Ç. Ö. Girit}
\email{caglar.girit@college-de-france.fr}
\affiliation{$\Phi_{0}$, JEIP, USR 3573 CNRS, Collège de France, PSL University, 11, place Marcelin Berthelot, 75231 Paris Cedex 05, France}

\date{\today}

%
%

\begin{abstract}
Superconducting circuits incorporating Josephson tunnel junctions are widely used for fundamental research as well as for applications in fields such as quantum information and magnetometry.
The quantum coherent nature of Josephson junctions makes them especially suitable for metrology applications.
Josephson junctions suffice to form two sides of the quantum metrology triangle, relating frequency to either voltage or current, but not its base, which directly links voltage to current.
We propose a five Josephson tunnel junction circuit in which simultaneous pumping of flux and charge results in quantized transconductance in units $4e^2/h = 2e/\Phi_0$, the ratio between the Cooper pair charge and the flux quantum.
The Josephson quantized Hall conductance device (JHD) is explained in terms of intertwined Cooper pair pumps driven by the AC Josephson effect.
We describe an experimental implementation of the device and discuss the optimal configuration of external parameters and possible sources of error.
The JHD has a rich topological structure and demonstrates that Josephson tunnel junctions are universal, capable of interrelating frequency, voltage, and current via fundamental constants.
\end{abstract}


\maketitle

\section{Introduction} 
\begin{figure}
\includegraphics[width=0.6\columnwidth]{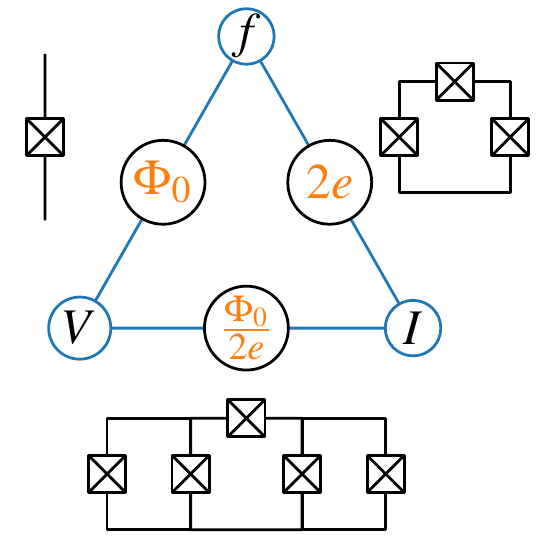}
\caption{\label{F1}\textbf{Completing the metrology triangle with only Josephson tunnel junctions.}
Circuits containing Josephson tunnel junctions (black crosses) can be used to form the metrology triangle relating voltage $V$, current $I$, and a pump signal at frequency $f$ via the fundamental constants $2e$, $\Phi_0=h/2e$, and their ratio, $R_Q = \Phi_0/2e = h/4e^2$.
In the AC Josephson effect (upper left), a microwave drive pumps flux quanta across a Josephson junction at a rate $f$ yielding the quantized voltage $V = \Phi_0 f$.
In a Cooper pair pump (upper right), the microwave drive pumps Cooper pairs (charge $2e$) at a rate $f$ yielding a quantized current $I = 2ef$.
A circuit with five Josephson junctions, the Josephson quantized Hall conductance device (bottom), combines both Cooper pair and flux pumping to yield a quantized Hall voltage $V_Y = R_Q I_X$ as in the quantum Hall effect.
}
\end{figure}

Among quantum coherent electronic components the most prominent are Josephson junctions and quantum Hall systems.
The physics describing electrons in both systems is rich and has yielded numerous applications in sensing, quantum information, and metrology.
Josephson junctions, due to their non-linearity, serve as the qubit building blocks of superconducting quantum computers~\cite{oliver_2020} and sensitive magnetometers~\cite{buchner_2018}.
In metrology this non-linearity, the Josephson relation, allows employing such junctions to define the voltage standard, with an accuracy much better than parts per billion~\cite{kautz_1996,rufenacht_impact_2018}.
Such junctions can also be used to obtain quantized currents, albeit with less accuracy than the Josephson voltage standard~\cite{pekola_single-electron_2013,kaneko_review_2016}.
Both metrological standards work by pumping Josephson junction circuits at a precise frequency $f$, obtaining either the quantized voltage $V = n\Phi_0f$ or the quantized current $I=2enf$, where $n$ is an integer and the fundamental constants are the magnetic flux quantum $\Phi_0 = h/2e$ and electron charge $e$.
In principle two sides of the quantum metrology triangle~\cite{poirier_ampere_2019},~\cref{F1}, which links frequency, voltage and current, can be completed using Josephson junctions only.
Given that the two non-commuting observables in a quantum circuit are number (charge) $\hat{N}$, and phase (flux) $\hat{\delta}$, it is not surprising that current and voltage, their respective time derivatives, can be quantized and used for metrology.

What is surprising however is that it has not been possible to use Josephson tunnel junctions to close the base of the metrology triangle, directly relating voltage to current.
Transconductance quantization is defined as a transverse or Hall voltage, $V_Y$, which is related to a longitudinal current $I_X$ by a resistance which depends only on fundamental constants.
Typical experimental implementations of transconductance quantization rely on semiconducting systems such as 2D electron gases in which there is a robust quantum Hall effect upon application of relatively large magnetic fields~\cite{poirier_qhe}.
With recent graphene based standards, it is possible to reduce the required magnetic field to several tesla and increase the operating temperature up to~\SI{10}{\kelvin}~\cite{ribeiro-palau_quantum_2015}.
In these semiconducting systems the relevant resistance is the von Klitzing constant $R_K = h/e^2$, and is quantized on the order of parts per billion.
For superconducting systems, the constant of proportionality between voltage and current would be the superconducting resistance quantum  $R_Q=h/4e^2$, which is more suggestively written as the ratio of the flux quantum to the Cooper pair charge $R_Q = \Phi_0/2e$.
This evocative relationship motivates the search for a Josephson junction circuit in which flux quanta and charge quanta are pumped simultaneously, producing a quantized resistance.

A circuit incorporating a Josephson tunnel junction and an $LC$ resonator was proposed to quantize transconductance, but requires an impractical quantum phase slip element~\cite{hriscu_2013}.
Non-trivial topology was identified in the Andreev bound state spectrum of multi-terminal superconducting devices~\cite{van_heck_single_2014,yokoyama_singularities_2015,xie_topological_2017} and it was shown that such systems could also exhibit a quantized Hall conductance~\cite{riwar_multi-terminal_2016,eriksson_topological_2017,meyer_non-trivial_2017,xie_weyl_2018}.
Although these multi-terminal weak link systems have motivated several experiments~\cite{draelos_supercurrent_2018,graziano_pribiag_2020,pankratova_multi-terminal_2020}, topological effects depend on the existence of highly transmitting microscopic Andreev states and device synthesis is challenging.
We propose a circuit containing only five Josephson tunnel junctions, the Josephson quantized Hall conductance Device (JHD), which quantizes $V_Y$ at sub-tesla magnetic fields while requiring only conventional fabrication techniques.
Not only is the Hamiltonian for our device completely different from that of Andreev based systems, engineering the circuit is relatively easy as the required technology is mature, there are less constraints on the circuit dimensions, no junction requires more than two terminals, and the junction transparencies may be small.

\section{Flux and Charge Quantization}

The upper sides of the quantum metrology triangle,~\cref{F1}, correspond to flux and charge pumping, processes which occur simultaneously in the JHD.
Flux pumping can be understood by considering the phase evolution in the AC Josephson effect.
A single Josephson junction,~\cref{F1} (upper left), biased at a voltage $V_J$ will have a superconducting phase $\delta$ which evolves linearly in time at the Josephson frequency $\omega_J = \dot{\delta} = V_J/\varphi_0$, where $\varphi_0 = \Phi_0/2\pi$ is the reduced flux quantum.
A $2\pi$ change in $\delta$ corresponds to pumping one fluxoid and occurs at a rate $f_J = \omega_J/2\pi$.
In Josephson voltage standards, a microwave signal at frequency $f$ is used to synchronize fluxoid pumping so that $f_J = nf$ and the voltage $V_J$ is determined with a precision limited only by the microwave reference clock and not by thermal noise in the DC voltage supply~\cite{kautz_1996}.
The topological nature of fluxoid quantization as well as charge quantization and the quantum Hall effect is highlighted by Thouless~\cite{thouless_1997,thouless_1998}.

\begin{figure*}
\includegraphics[width=0.9\textwidth]{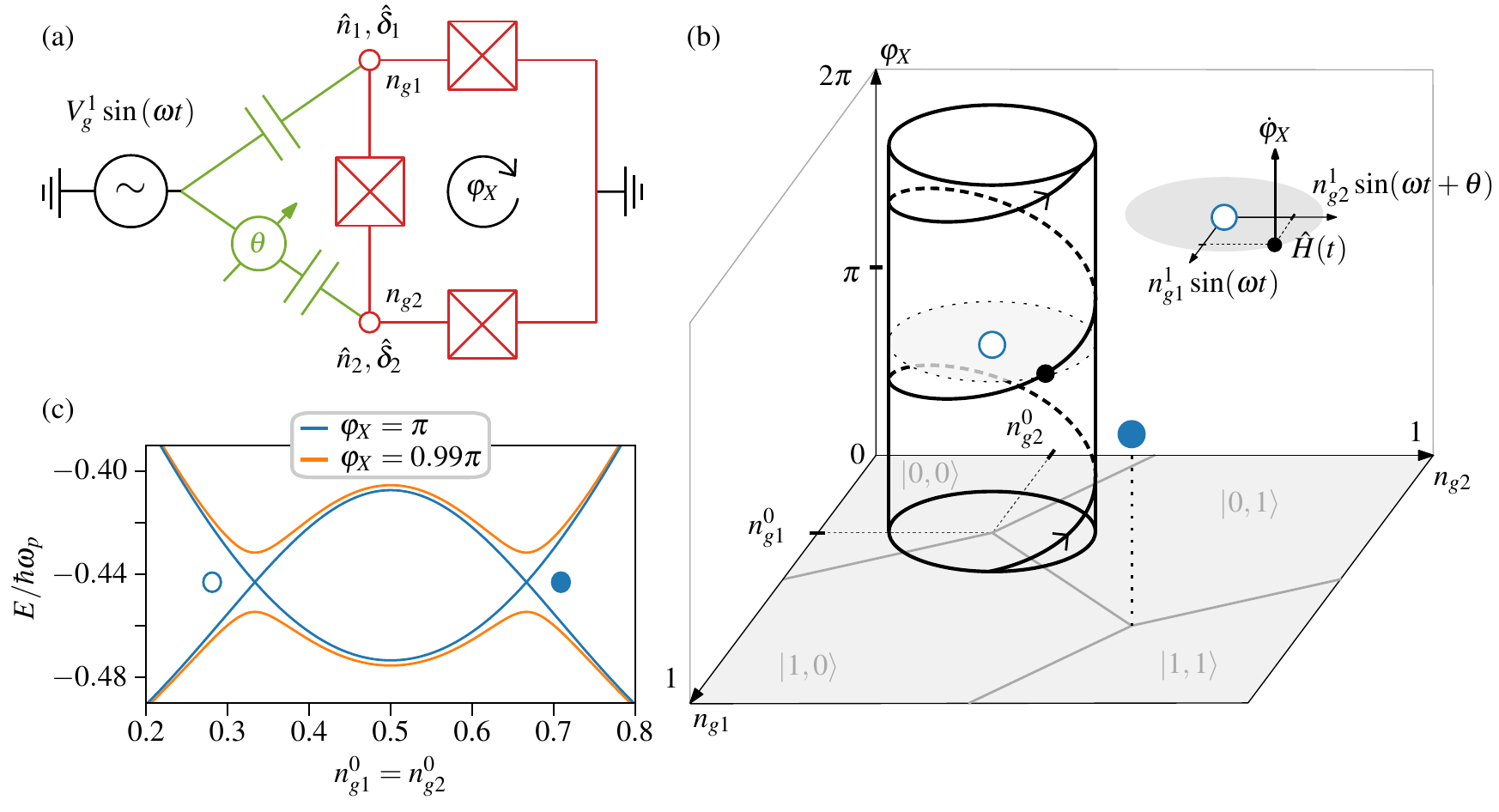}
\caption{\label{F2}\textbf{Topological Cooper pair pumping}
(a) The Cooper pair pump (CPP) consists of three Josephson tunnel junctions in series (red boxed crosses), forming two superconducting islands with canonical quantum variables $\hat{n}_{1,2}$ and $\hat{\delta}_{1,2}$.
Gate voltages applied via a microwave pump and additional DC sources (not shown) determine the charge offsets $n_{g1},n_{g2}$ on the islands.
An external magnetic field tunes the reduced magnetic flux $\varphi_X$.
(b) The three periodic parameters $n_{g1},n_{g2}$ and $\varphi_X$ form a parameter space analogous to a 3D Brillouin zone~\cite{leone_topological_2008}.
Degeneracies in the spectrum are indicated by blue dots on the $\varphi_X=\pi$ plane and are associated with topological charges $\pm1$.
The current flowing through the CPP, $I = 2ef$, is quantized on a cyclic trajectory in parameter space which encloses a degeneracy ($f = \omega/2\pi$).
(c) The two lowest energy bands of the circuit are plotted for equal DC charge offset $n_{g1}^{0} = n_{g2}^{0}$ and $E_J/E_C=1$.
There are two degeneracies in the spectrum for $\varphi_X=\pi$ (blue) and none for other values (e.g. $\varphi_X=0.99\pi$, red).
Energy is plotted in units of plasma frequency $\hbar\omega_p = \sqrt{2E_J E_C}$ and all junctions are identical.}
\end{figure*}

To relate frequency to current,~\cref{F1} (upper right), the relevant superconducting circuit is the Cooper pair pump (CPP),~\cref{F2}(a), consisting of three Josephson tunnel junctions (red boxed crosses) in series, forming two superconducting islands with canonical quantum variables $\hat{n}_{1,2}$ and $\hat{\delta}_{1,2}$~\cite{geerligs_single_1991,erdman_fast_2019}.
For simplicity we consider identical junctions.
We define the charging energy as $E_C=(2e)^2/2C$, where $C$ is the junction capacitance.
Charge offsets $n_{g1},n_{g2}$ have DC components $n_{g1}^0,n_{g2}^0$ determined by static gate biases (not shown) and AC components $n_{g1}^1,n_{g2}^1$ determined by a microwave pump of amplitude $V_g^{1}$ and radial frequency $\omega$.
A $\theta$-phase shifter (green) allows dephasing the oscillating parts on each island.

Ignoring the Josephson part of the Hamiltonian, the CPP has stable charge states on hexagonal plaquettes shown in the bottom plane of~\cref{F2}(b), delineated by gray lines denoting charge degeneracies.
The AC modulation of both gate voltages around a plaquette vertex, a point of triple degeneracy, can be used to cycle between charge states and drive exactly one Cooper pair across the device per cycle.
The Josephson coupling terms hybridize the charge states with strength $E_J$, the Josephson energy.
The characteristic energy scale is now $\hbar\omega_p = \sqrt{2E_J E_C}$, where $\omega_p=1/\sqrt{L_JC}$ is the plasma frequency, and the dimensionless impedance $\alpha = Z_J/R_Q = 1/2\pi\cdot\sqrt{2E_C/E_J}$ characterizes the ratio between charging energy and Josephson energy.
The Josephson inductance is defined by $E_J = \varphi_0^2/L_J$ and is related to the critical current $I_0 = \varphi_0/L_J$.
The reduced magnetic flux $\varphi_X=B_XA/\varphi_0$, determined by an external magnetic field $B_X$ threading the three-junction loop of area $A$, can be used to change the Josephson coupling between islands.
Non-trivial topological effects can arise when the energy spectrum has degeneracies at certain points in the parameter space.
At $\varphi_X=\pi$ the Josephson coupling is effectively turned off, so there are two degeneracies,~\cref{F2}(c), and modulating the gate voltages pumps exactly one Cooper pair each time the trajectory winds around a degeneracy.
But for any other value of $\varphi_X$, the degeneracies are lifted and a gate voltage cycle no longer results in quantized charge transfer.

This error in pumped charge can in principle be averaged out by covering a closed surface around the degeneracy with a helical trajectory as shown in~\cref{F2}(b)~\cite{leone_cooper_2008,erdman_fast_2019}.
The helix maps out a cylinder-like surface centered at charge offset $n_{g1}^0,n_{g2}^0$.
The radial profile is determined by the AC amplitude $V_{g}^{1}$ and phase shift $\theta$ whereas the upward velocity is given by $\dot{\varphi}_X$. 
Taking into account the $2\pi$ periodicity of $\varphi_X$, the cylinder shown in~\cref{F2}(b) ($\theta = \pi/2$) is actually a torus $T^2$ in parameter space.
The pumped charge is proportional to the integral of the Berry curvature over $T^2$ which is equal to $2\pi$ times the Chern number $C(T^2)$.
The average current across the CPP is then given by $I=2ef C(T^2)$, where the microwave pump frequency $f = \omega/2\pi$ is also the winding rate in the $n_{g1},n_{g2}$ plane.
It is interesting to note that this current does not depend on the value nor on the sign of $\dot{\varphi}_X$ as long as $\dot{\varphi}_X \ne 0$ and is incommensurate with $\omega$~\cite{leone_cooper_2008}.
The phase ramp $\dot{\varphi}_X$ can be applied by inductively coupling to the CPP loop~\cite{leone_cooper_2008} or inserting a voltage source~\cite{geerligs_single_1991}.

Current quantization is insensitive to small variations in junction critical currents and capacitances as the resulting modifications to $E_{Ji},E_{Ci}$ only move the degeneracies in the $\varphi_X=\pi$ plane but do not destroy them.
The first experiments employing topological pumping of CPPs had low currents and significant error~\cite{geerligs_single_1991}, but optimization can mitigate factors such as non-adiabaticity and supercurrent leakage~\cite{leone_cooper_2008}, resulting in improved performance~\cite{erdman_fast_2019}.
Significant amelioration is still necessary before they can serve as current standards for metrology~\cite{kaneko_review_2016}, and larger currents are obtained by combining the conventional QHE and Josephson voltage standard~\cite{brun-picard_2016}.

\begin{figure}
\includegraphics[width=0.9\columnwidth]{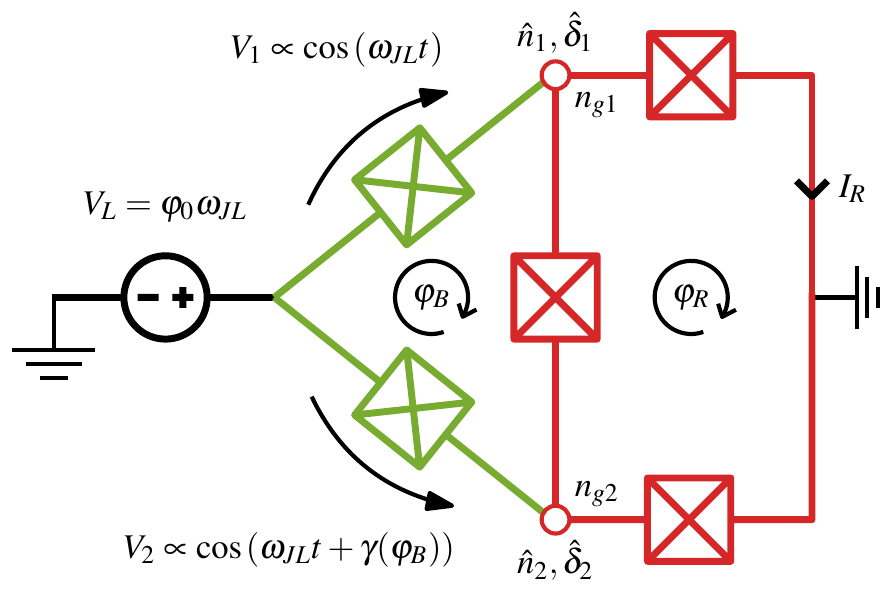}
\caption{\label{F3}\textbf{Driving a Cooper pair pump with the AC Josephson effect}
  The constant voltage bias $V_L$ is converted by the green Josephson junctions (boxed crosses) into oscillating currents $I_L\approx I_0\sin(\omega_{JL}t)$, where $I_0$ is their critical current.
  These currents generate oscillating voltages $V_1$ and $V_2$ at the charge nodes of a Cooper pair pump (CPP) formed by the three red junctions.
  The reduced magnetic flux $\varphi_B=\Phi_B/\varphi_0$ threading the central loop introduces a phase offset $\gamma(\varphi_B)$ between these oscillating voltages ($\varphi_B=\Phi_0/2\pi$).
  An additional ramp of the flux $\varphi_R$ results in a quantized current $I_R = 2e\omega_{JL}/2\pi = 4e^2/h \cdot V_L$ flowing through the red CPP.
}
\end{figure}

\section{Transconductance quantization}

A circuit exploiting the AC Josephson effect to replace both microwave pumps of the CPP by voltage-biased Josephson junctions allows combining flux and charge pumping to directly link voltage and current via a quantized conductance.
The key point is that the microwave pump frequency $f$ can be associated with a Josephson frequency $\omega_J = 2\pi f$ and the pump amplitude $V_{g}^{1}$ with the critical current.
The five junction circuit shown in~\cref{F3} is one realization of a Josephson quantized Hall conductance device which closely resembles the CPP circuit in~\cref{F2}(a).
Due to the Josephson effects the green Josephson junctions convert the input DC voltage $V_L$ into oscillating currents at frequency $\omega_{JL}$ and amplitude proportional to their critical currents $I_0$.
We assume that the critical currents of the other junctions (red) are large enough such that the DC voltage $V_L$ drops only across the green junctions.
Although this assumption is not essential for transconductance quantization, it helps establishes the analogy with the CPP microwave driving circuit in~\cref{F2}(a).
The oscillating Josephson currents result in oscillating voltages $V_1$ and $V_2$ which drive the charge nodes of the CPP formed by the three red junctions on the right.
Due to the topologically non-trivial nature of the CPP spectrum, current quantization holds for a large range of amplitudes for $V_1,V_2$ determined by the junction impedances $Z_{Ji} = \sqrt{L_{Ji}/C_{Ji}}$.
The phase difference $\gamma(\varphi_B)$ between the two leftmost green junctions can be tuned with the reduced magnetic flux $\varphi_B$, with the function $\gamma$ accounting for the current-phase dependence of the three-junction loop~\cite{zapata_1996,sterck_2005}.
Compared to the microwave gate drives of the CPP of~\cref{F1}(a), the amplitudes $V_{1,2}$ correspond to $V_{g}^{1}$ and $\gamma(\varphi_B)$ to the phase shift $\theta$.
With additional DC gate biasing to obtain the proper charge offsets on the superconducting islands and a series voltage source $V_R$ in the red CPP loop to ramp $\varphi_R$, we completely reproduce the pumping protocol of~\cref{F2}.
The current pumped in the right CPP is quantized and given by $I_R=2e\omega _{JL}/2\pi = V_L\cdot 4e^2/h = V_L/R_Q$.

\begin{figure*}
\includegraphics[width=0.85\textwidth]{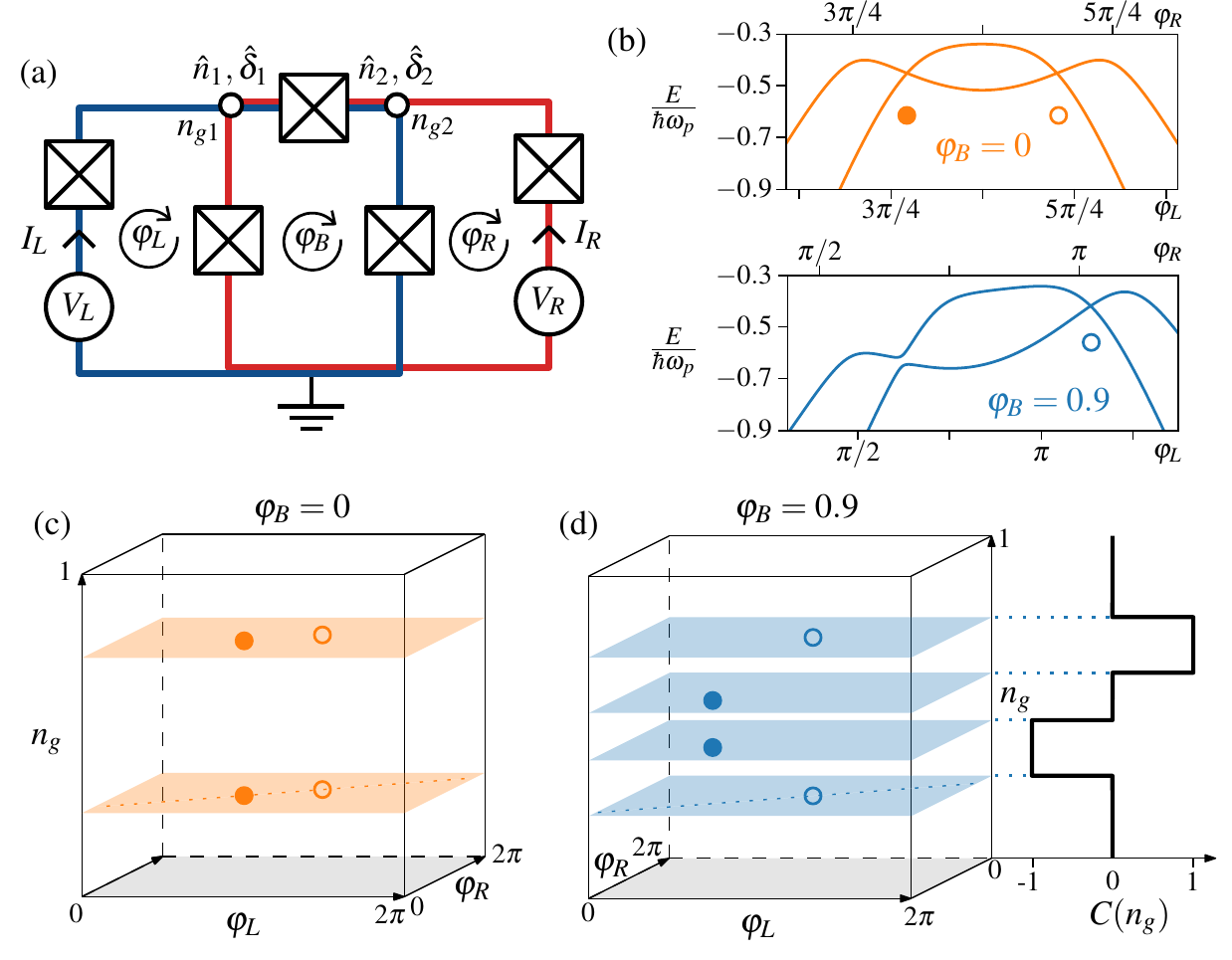}
\caption{\label{F4}\textbf{Topological properties of the Josephson quantized Hall conductance device}
  (a) The circuit consists of five Josephson tunnel junctions (boxed crosses) forming two superconducting islands (circled) and three loops.
  Gate voltages applied to each island (not shown) determine the global offset charge $n_g$ and along with the loop fluxes $\varphi_{L,B,R}$ tune the energy spectrum of the system.
  The constant voltage sources $V_L$ and $V_R$ allow linearly ramping $\varphi_L$ and $\varphi_R$ and supply currents $I_{L,R}$.
  (b) The two lowest energy bands containing degeneracies are plotted for $n_g \approx 0.25$ along cuts indicated in the bottom $\varphi_L,\varphi_R$ planes of (c) and (d) (dashed lines).
  As $\varphi_B$ is tuned away from zero (top, orange) one of the degeneracies is lifted (bottom, blue).
  Degeneracies with topological charge $+1$ ($-1$) are indicated by filled (unfilled) circles.
  (c) The positions of degeneracies are indicated in the 3D parameter space $\varphi_L$, $\varphi_R, n_g$.
  For $\varphi_B=0$ pairs of degeneracies with opposite signs are located on planes of constant $n_g$, resulting in a Chern number of zero.
  (d) For non-zero $\varphi_B$ the $+1$ degeneracies (filled circles) split off the planes in (c) and shift toward $n_g = 0.5$.
  The Chern number, plotted on the right, is $\pm1$ for $n_g$ lying between two opposite charge degeneracies and zero elsewhere.
  Details of the Hamiltonian, parameters, and positions of the degeneracies are provided in the~\cref{A},~\cref{FA1}.
}
\end{figure*}

Although the JHD shown in~\cref{F3} is conceptually closest to the CPP of~\cref{F2}(a), the symmetric circuit of~\cref{F4}(a) also quantizes transconductance and more clearly demonstrates that there are actually two intertwined Cooper pair pumps, indicated in blue and red.
One JHD circuit can be transformed into another by shifting the sources and rotating the branches (see \cref{A}~\cref{FA2}).
In the following we show by topological arguments that each time a flux quantum is pumped in one CPP loop of the symmetric JHD, a Cooper pair is pumped in the other CPP loop, resulting in transconductance quantization.

For the numerically computed spectra and degeneracies in~\cref{F4}(b-d) we assume for simplicity that the charge offset on both islands is $n_g$.
To show that symmetry is not necessary for transconductance quantization we treat the experimentally relevant situation in which the plasma frequency of all junctions is identical but not their surface areas.
The full Hamiltonian, junction parameters, positions and topological charges of the degeneracies, as well as a description of numerical methods, are provided in the \cref{A} (\cref{FA1}).

The Josephson quantized Hall conductance device, which realizes the quantum Hall effect with only Josephson tunnel junctions, can be understood by examining the topological properties of the circuit.
Transconductance quantization can be linked to the system's Hamiltonian and eigenstates via a Chern number (see \cref{A}).
As with the degeneracies in the energy spectrum of the CPP those of the JHD,~\cref{F4}(b), are also associated with topological charges $\pm 1$ (filled and unfilled circles, respectively).
The Brillouin zone $\varphi_L,\varphi_R,n_g$ and positions of degeneracies are shown in~\cref{F4}(c-d).
As in the Andreev state based topologically non-trivial systems~\cite{riwar_multi-terminal_2016,eriksson_topological_2017,meyer_non-trivial_2017} to encompass a degeneracy the $\varphi_L,\varphi_R$ plane is swept by applying constant voltages $V_L = \varphi_0\dot{\varphi}_L$ and $V_R = \varphi_0\dot{\varphi}_R$.
Unlike the multi-terminal Andreev devices which have a different Hamiltonian, the JHD has charge offset parameters $n_g$ in addition to fluxes $\varphi_{B,L,R}$.
When a plane crosses a degeneracy the corresponding Chern number $C(n_g,\varphi_B)$ changes by the value of the topological charge.
We take the convention that topological charges are added to the Chern number in the direction of increasing $n_g$.
For $\varphi_B=0$, shown in orange in~\cref{F4}(b,c), pairs of degeneracies with opposite topological charges are located on the same $n_g$ plane and the Chern number is always zero.
On the contrary for non-zero values such as $\varphi_B = 0.9$, shown in blue in~\cref{F4}(b,d), single degeneracies exist in the  $\varphi_L,\varphi_R$ plane for certain values of $n_g$ and the Chern number can be $\pm1$.

For a given eigenstate the associated instantaneous current through one of the loops has a contribution from the junction supercurrents, but this dynamical term averages out to zero for an adiabatic sweep of the entire $\varphi_L,\varphi_R$ plane.
The second contribution comes from the Berry curvature which when integrated over this plane is proportional to the Chern number.
This geometric contribution gives rise to DC currents $I_L$ ($I_R$) which are quantized~\cite{gritsev_dynamical_2012,riwar_multi-terminal_2016},
\begin{equation}
  I_{L,R}(n_g,\varphi_B) = \frac{4e^2}{h} C(n_g,\varphi_B) V_{R,L},
  \label{E1}
\end{equation}
and depend on the voltage $V_R$ ($V_L$) applied to the opposite loop (see \cref{A}).
Referring to the circuit~\cref{F4}(a), this transconductance is interpreted as two Cooper pair pumps (red and blue loops) acting as AC Josephson drives for one another such that the pumped current through one depends on the Josephson frequency of the other.
The current circulating in the middle loop threaded by $\varphi_B$, unlike $I_L$ and $I_R$, is not quantized.
Although the eigenstates of the multi-terminal Andreev systems are not the same as for the JHD, since \cref{E1} is independent of the basis, transconductance is quantized in both systems.

\section{Discussion}

The topology of the JHD and isolated CPP can be compared to understand why flux and charge are pumped in the JHD and only charge is pumped in the CPP.
The cylinder of~\cref{F2}(b) and the plane of~\cref{F4}(c-d) both correspond to tori covering only one degeneracy and imply charge pumping.
The CPP trajectory in the $n_{g1},n_{g2}$ plane does not reach the Brillouin zone boundaries, unlike the vertical component ($\varphi_X$).
On the contrary the JHD bias voltages $V_{L,R}$ result in both parameters $\varphi_L,\varphi_R$ crossing the Brillouin zone boundary.
These crossings correspond to pumping of two fluxoids in the JHD circuit loops.

\begin{figure*}
\includegraphics[width=0.7\textwidth]{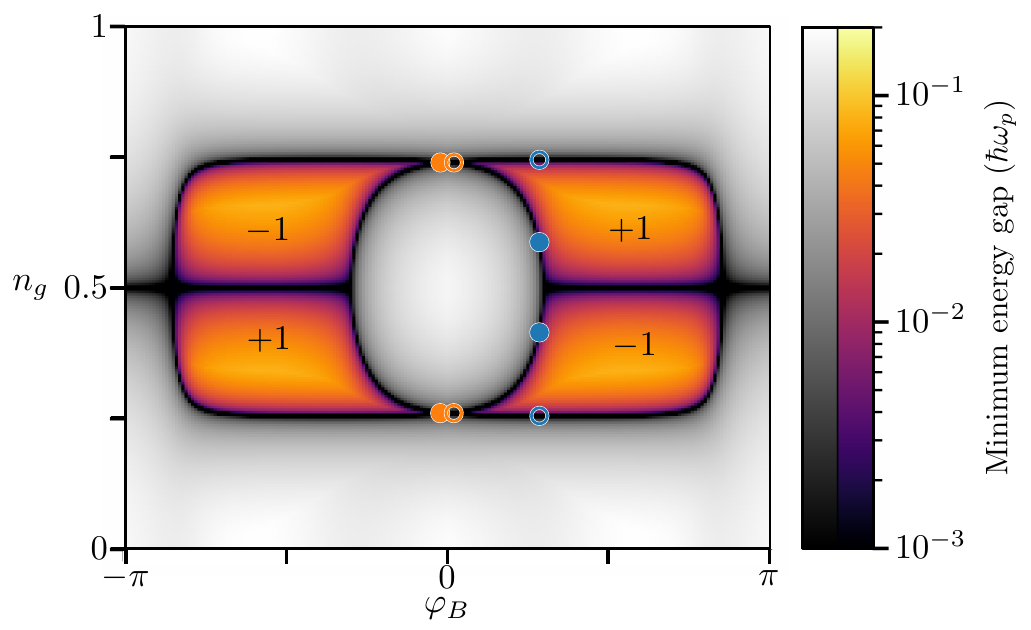}%
\caption{\label{F5}\textbf{Optimizing external parameters of the Josephson quantized Hall conductance device.}
  The minimum energy gap for the symmetric Josephson quantized Hall conductance device is plotted as a function of global charge offset $n_g$ and loop flux $\varphi_B$.
  Energies are scaled to the plasma frequency $\hbar\omega_p$, with a lower cutoff at $10^{-3}$, and determined by minimizing the energy gap as a function of $\varphi_L,\varphi_R$ for each value of $n_g$ and $\varphi_B$.
  The topological charge of degeneracies is indicated by filled ($+1$) and unfilled ($-1$) circles and correspond to the degeneracies in~\cref{F4}(b-d).
  The Chern number $C(n_g,\varphi_B)=\pm1$ is indicated in four topologically non-trivial regions delineated by black lines where the gap vanishes.
  Elsewhere the system is topologically trivial (zero Chern number, gray regions).
  Fixing $n_g$ and $\varphi_B$ to have a large energy gap in a topologically non-trivial region results in transconductance quantization with minimal error due to non-adiabatic transitions.
}
\end{figure*}

The value of the Chern number $C(n_g,\varphi_B) = \pm1$ as well as the minimum energy gap in the $\varphi_L,\varphi_R$ plane is plotted in the phase diagram~\cref{F5} as a function of $n_g$ and $\varphi_B$ for the same junction configuration $E_{Ji},E_{Ci}$ as in~\cref{F4}.
The dark blue lines where the gap vanishes demarcate the different topological regions.
The Chern number in the central and exterior trivial regions is equal to zero and its label is omitted.
The relative placement of Chern numbers can be explained by general symmetry arguments for Josephson junction circuits~\cite{valla_2020}.
Overall inversion symmetry of the Hamiltonian implies that the Chern number is conserved when all parameters are inverted.
This results in the configuration of~\cref{F5}, where inverting both $n_g$ and $\varphi_B$, and implicitly $\varphi_L$ and $\varphi_R$, maintains the sign of the Chern number.
Time reversal symmetry on the other hand corresponds to inverting either all charge parameters or all flux parameters, changing the sign of the Chern number.
As a result the topologically non-trivial region has a quadrupole like distribution which occupies a large fraction of the phase space.
The size of this region will shrink as junction disorder increases, but as long as the operating point for the JHD remains in a region of non-zero Chern number, the pumped current will be quantized.

In addition to labels for the Chern number,~\cref{F5} has blue and orange circles which correspond to the degeneracies of~\cref{F4}(b-d).
For a charge of a given sign (filled or unfilled circle), continuity implies that it follows the dark blue degeneracy lines as $n_g$ and $\varphi_B$ are varied.
For example the $-1$ charge at $n_g \approx 1/4; \varphi_B = 0$ (bottom unfilled orange circle) moves horizontally to the right whereas the $+1$ charge (bottom filled orange circle) moves upwards towards $n_g = 1/2$, flanking the non-trivial region in the lower right with Chern number $-1$.
The two $+1$ charges (filled blue circles) join at the line $n_g = 0.5$ as $\varphi_B$ increases, eventually followed by the two $-1$ charges (unfilled blue circles) as $\varphi_B$ approaches $\pi$.
When all junctions are identical, the outer corners of the four non-trivial regions extend out to $n_g \approx 1/4, 3/4;\; \varphi_B=\pm\pi$ and the size of the central trivial region shrinks as the ratio $E_J/E_C$ increases (\cref{A}~\cref{FA3}).

Transconductance quantization only holds in the adiabatic limit, so it is important that the pump frequencies, determined by $V_L$ and $V_R$, are small compared to the plasma frequency~\cite{leone_topological_2008}.
This limitation was studied in detail for Andreev multi-terminal devices in which the corresponding energy scale, the superconducting gap, is several times larger~\cite{riwar_multi-terminal_2016,eriksson_topological_2017}.
A typical value for the plasma frequency of aluminum Josephson junctions is \SI{20}{GHz}, corresponding to $\varphi_0\omega_p =$~\SI{40}{\micro\volt}.
Applied voltages must be much smaller than $\varphi_0\omega_p$ to avoid inducing Landau-Zener transitions (LZT) from the ground state to excited states.
As the LZT probability scales inversely with the square of the energy gap, the JHD should be operated at values of $n_g, \varphi_B$ that maximize the smallest gap along the trajectory, the $\varphi_L,\varphi_R$ plane.
From~\cref{F5} this optimum is near $n_g=1/3$ and $\varphi_B=\pi/2$, where the minimum gap is roughly equal to $0.1\omega_p$ or \SI{2}{GHz} (\SI{4}{\micro\volt}).
For good quantization the applied voltages must be smaller than this value, resulting in pumped currents which are smaller than~\SI{150}{\pico\ampere}.

In the case of identical junctions the choice of the junction impedance $\alpha = Z_J/R_Q$, requiring careful device design and junction fabrication, is also important to minimize noise.
The parameter $\alpha$ is an effective fine structure constant and is inversely proportional to the junction surface area.
By making small junctions one can obtain much larger values than $\alpha_0 = Z_0/8R_Q \approx 1/137$, where $Z_0$ is the vacuum impedance, greatly modifying the phase diagram of~\cref{F5}.
Comparison is made for different values of $\alpha$ in \cref{A}~\cref{FA3}.
We find that the minimum gap is largest for $\alpha \simeq 1/2$, allowing increased pump currents while remaining adiabatic.
Optimizing the value of $\alpha$ for transconductance quantization is analogous to impedance matching $Z_J$, in this case to approximately $R_Q/2$.
Although deviating from the optimum will decrease the gap, reducing $\alpha$ also comes at the cost of enhanced supercurrent leakage, as noise in $\varphi_B$ and $n_g$ prevents the dynamical contribution of the current from averaging to zero during the $\varphi_L,\varphi_R$ sweep.
Other sources of error which have already been analyzed in depth for Cooper pair pumps~\cite{leone_cooper_2008,erdman_fast_2019} are the effect of the external biasing circuit, thermal population of excited states, co-tunneling, and quasiparticle poisoning.
Although some proposed error mitigation techniques can be directly applied to our circuit, implementing other techniques such as shortcut-to-adiabaticity~\cite{muga_sta} would be more difficult.

Fabricating the Josephson quantized Hall conductance device, given current technology for superconducting circuits containing hundreds of Josephson tunnel junctions, is straightforward.
This is in contrast to quantized Hall transconductance devices based on phase-slip or high-transparency multi-terminal weak links.
Using niobium, with a higher critical current density than aluminum, is desirable for increasing the plasma frequency and extending the range for adiabatic operation.
Static control of loop fluxes can be accomplished with inductive coupling and charge offsets can be adjusted with local gates.

As with conventional quantum Hall resistance standards, and unlike Josephson current or voltage standards, the Josephson quantized Hall conductance device does not require an external microwave pump.
Applying voltages $V_L$ and $V_R$ could exploit strategies from CPP experiments such as inserting small resistances into the left and right loops which are biased by external current sources~\cite{vartiainen_nanoampere_2007}.
A more clever strategy would directly apply a voltage difference across these two resistances.
To preserve phase coherence over time scales comparable to the pump frequency, the resistances $r$ should be small enough such that $I_0r \ll V_{L,R}$~\cite{aunola_connecting_2003}. 
Although noise of the external sources should be minimized in principle the pumped current will follow fluctuations in $V_L$ and $V_R$ such that the transconductance remains quantized.
Measurement of the pumped currents $I_L$ and $I_R$ can be made by borrowing techniques from CPP experiments~\cite{mottonen_experimental_2008,erdman_fast_2019}.
One possibility is using a SQUID current amplifier, possibly combined with a cryogenic current comparator, which would have high sensitivity but requires inserting an inductor in the CPP loops~\cite{brun-picard_2016}.
A more detailed error analysis, including the impact of the biasing and measurement scheme, as well as consideration of niobium Josephson junctions for larger pump currents, is needed for a complete evaluation of the circuit for a possible resistance standard.

While the Josephson quantized Hall conductance device is unlikely to compete with existing metrological resistance standards, establishing experimentally that it quantizes transconductance is of immediate interest.
The first step would be to verify the degeneracy structure with microwave spectroscopy in a superconducting circuit QED geometry and obtain an experimental equivalent of~\cref{F5}.
Such spectroscopy measurements have been performed for superconducting qubits with drive-induced topology~\cite{leek_observation_2007,roushan_observation_2014,schroer_measuring_2014}.
It may be possible to directly measure the local topological properties with such experiments~\cite{ozawa_extracting_2018,klees_microwave_2018,tan_experimental_2019}.
If direct measurement of the transconductance encounters problems due to noise from the biasing circuit, techniques which avoid DC connections may be possible~\cite{nguyen_current_2007}.
On the other hand one could keep the DC connections and use microwave pumps to synchronize the DC voltages $V_L$ and $V_R$, significantly reducing noise.

We have observed that a circuit with an additional Josephson junction, giving it tetrahedral symmetry, also quantizes transconductance.
This \emph{tetrahedron} was considered previously as a candidate for a protected qubit~\cite{feigelman_superconducting_2004}.
It also has a rich topological structure and is part of the class of Weyl Josephson circuits~\cite{valla_2020}.
From heuristic arguments and numerical calculations we conjecture that Josephson circuits with fewer than five tunnel junctions cannot quantize transconductance.
However such circuits may still display topologically non-trivial phenomena in their spectra, including the possibility of merging Dirac points~\cite{monjou_merging_2013,tan_realizing_2017,tan_topological_2018}.
Another variant five-junction circuit is the dual of the JHD, \cref{A}~\cref{FA2}(a), a diamond shaped circuit with three charge nodes and two loops which may also quantize transconductance.

Many other theoretical questions remain including a rigorous validation of the simultaneous fluxoid-charge pumping mechanism,
determining the precise relationship between transconductance in JHD and the Andreev multi-terminal systems, and computing additional topological invariants~\cite{zhang_four-dimensional_2001,fan_kane_2014,palumbo_tensor_2019,tan_experimental_2020}.
Establishing the theoretical optimum values for parameters such as $\alpha$ so as to maximize the energy gap and minimize errors is necessary and implies a deeper understanding of the behavior of the phase diagram~\cref{F5}. 
Investigating the topological properties of arbitrary Josephson Hamiltonians~\cite{valla_2020} may lead to their general classification and open possibilities for novel applications of quantum circuits.
Our work shows that Josephson tunnel junctions are universal in the sense that they can connect the three sides of the quantum metrology triangle relating $f,V$ and $I$.
A major question remains as to whether such circuits can exhibit topological effects which go beyond this triangle.

\begin{acknowledgments}
  We thank Valla Fatemi, Landry Bretheau, and Anton Akhmerov for insightful comments and corrections.
  We also thank Julia Meyer, Benoît Douçot, Daniel Estève, Hugues Pothier, and Fabien Lafont for a critical reading of our manuscript.
  We acknowledge support from Jeunes Equipes de l'Institut de Physique du Collège de France.
  This research was supported by IDEX grant ANR-10-IDEX-0001-02 PSL and a Paris ``Programme Emergence(s)'' Grant.
  This project has received funding from the European Research Council (ERC) under the European Union's Horizon 2020 research and innovation programme (grant agreement 636744).
\end{acknowledgments}

\appendix*

\section{\label{A}}



\begin{figure}
\includegraphics[width=\columnwidth]{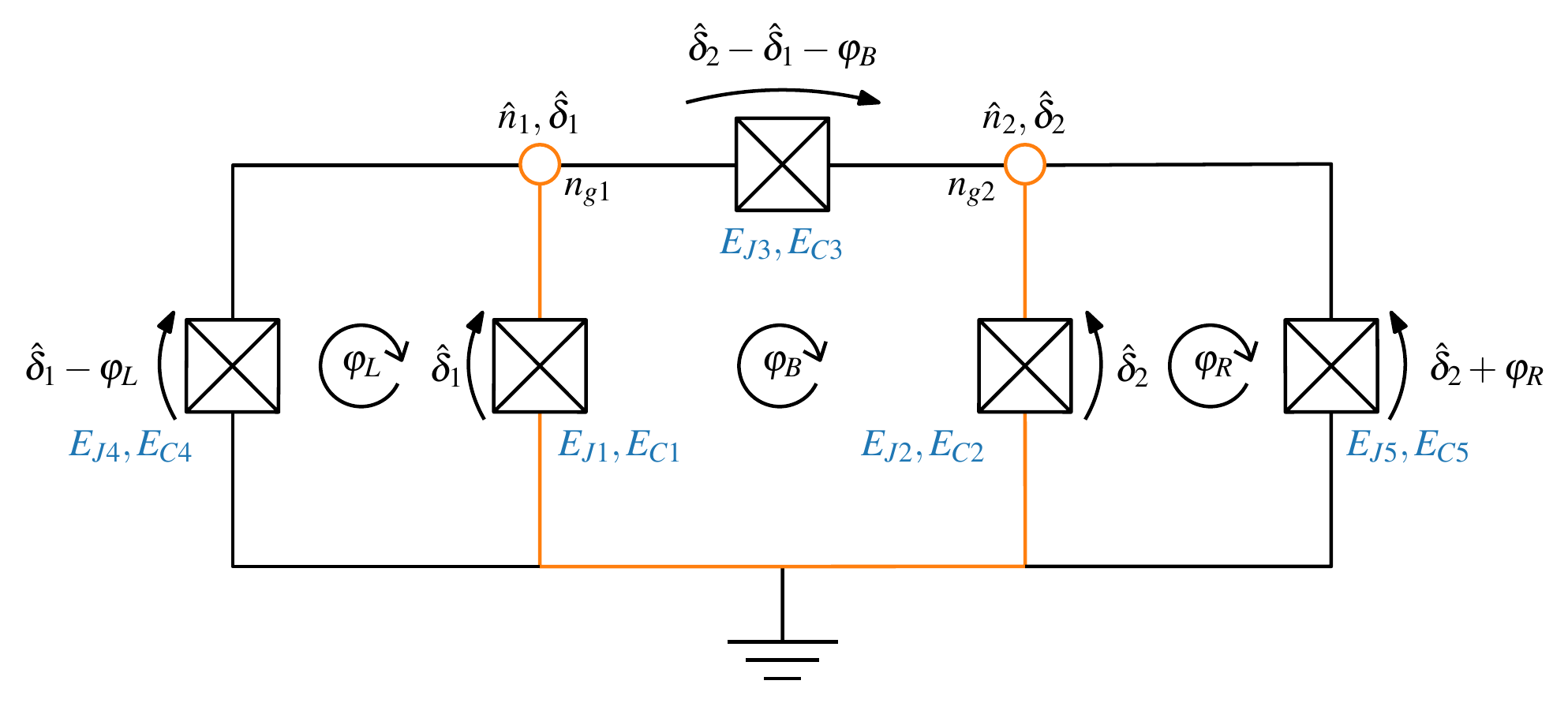}
\caption{\label{FA1}\textbf{Full circuit of symmetric Josephson quantized Hall conductance device.}}
\end{figure}

The Hamiltonian $H_J + H_C$ for the symmetric JHD circuit of the main text, reproduced in~\cref{FA1} with all parameters identified is
\begin{gather*}
H_J=-E_{J1}\cos\hat{\delta}_1-E_{J2}\cos\hat{\delta}_2-E_{J3}\cos(\hat{\delta}_2-\hat{\delta}_1-\varphi_B) \\ -E_{J4}\cos(\hat{\delta}_1-\varphi_L)-E_{J5}\cos(\hat{\delta}_2+\varphi_R), \\
  H_C = 2e^2 (\mathbf{\hat{n}} -\mathbf{n_g}) ^T C^{-1} (\mathbf{\hat{n}}-\mathbf{n_g}),
\end{gather*}
where the charge operators are $\mathbf{\hat{n}} = (\hat{n}_1,\hat{n}_2)$, the charge offsets are $\mathbf{n_g} = (n_{g1},n_{g2})$, and the capacitance matrix $C$ is given by
\begin{equation*}
C=\begin{pmatrix}
C_{J1}+C_{J3}+C_{J4} & -C_{J3} \\
-C_{J3}& C_{J2}+C_{J3}+C_{J5}\\
\end{pmatrix}.
\end{equation*}

The gate capacitances, in general small compared to $C_{Ji}$, are neglected in $C$.
The individual charging energies are $E_{Ci}=2e^2/C_{Ji}$.

The Chern number associated to a 2D plane spanning parameters $X$ and $Y$ can be computed from the Berry curvature of the ground state $\ket{\psi}$
\begin{equation*}
B_{X,Y}= -2 \text{Im}\Braket{\frac{\partial \psi}{\partial X}| \frac{\partial \psi}{\partial Y} }.
\end{equation*}
For a $\varphi_L,\varphi_R$ sweep and a given set of equal charge offsets $n_g$ and reduced flux $\varphi_B$, we define the Chern number as the integral of $B_{\varphi_R,\varphi_L}$ over the whole $\varphi_L,\varphi_R$ plane:
\begin{equation*}
C(n_g,\varphi_B)=\frac{1}{2\pi}\int_0^{2\pi}\int_0^{2\pi} d\varphi_L d\varphi_R B_{\varphi_R,\varphi_L}.
\end{equation*}
Following~\cite{gritsev_dynamical_2012,riwar_multi-terminal_2016} we obtain~\cref{E1}, where we have a plus sign for both $I_L$ and $I_R$ since positive voltage $V_R$ corresponds to negative $\dot{\varphi}_R$ given the circuit conventions of~\cref{FA1}.

For numerical calculations, circuit Hamiltonians are directly written in the charge basis with typically ten charge states for each island.
Eigenvalues and eigenstates are obtained by direct diagonalization of the sparse Hamiltonian matrix via the Lanczos algorithm as implemented in~\href{https://www.scipy.org/}{scipy}.
To determine the Berry curvature the gradient of the Hamiltonian with respect to external parameters is calculated analytically and then converted to the charge basis.
Chern numbers are obtained by numerical integration of the Berry curvature over the desired 2D surface in parameter space.
The precise locations of degeneracies are obtained with minimization techniques such as simplicial homology global optimization in~\href{https://docs.scipy.org/doc/scipy/reference/generated/scipy.optimize.shgo.html}{scipy}.

Parameters used to obtain the spectra and degeneracies in~\cref{F4} are $E_{J1}=1.0, E_{J2}=0.8, E_{J3} = 1.1, E_{J4}=0.9, E_{J5} = 1.2$ and we keep the plasma energy constant $\hbar\omega_p = \sqrt{2E_{Ji}E_{Ci}} = 1$ such that $E_{Ci} = 1/E_{Ji}$.
This corresponds to the experimentally relevant situation where the surface area of the Josephson junctions may be different but since the oxidation process for the tunnel barriers is common, the plasma frequencies are the same.

For the spectra of~\cref{F4}(b) the cuts are made in the $\varphi_L,\varphi_R$ plane at fixed $\phi_B,n_g$ along the diagonals shown in the bottom planes of~\cref{F4}(c,d) and given by the following equations,
\begin{align*}
  \varphi_R &= 0.7195\varphi_L+0.8811 \; (\varphi_B = 0, n_g = 0.2598,\textrm{orange}), \\
  \varphi_R &= 0.7195\varphi_L+0.7092 \; (\varphi_B = 0.9, n_g = 0.2541,\textrm{blue}).
\end{align*}

The positions and topological charges of degeneracies in~\cref{F4}(c,d) are given in~\cref{AT1}.
The source code is available on Zenodo~\cite{peyruchat_zenodo_2020}.

\begin{table*}
\caption{\label{AT1}Positions in parameter space $\varphi_L, \varphi_R, n_g$ and topological charges $\chi$ of degeneracies in~\cref{F4}}
\begin{ruledtabular}
  \begin{tabular}{cccc|cccc}
    \multicolumn{4}{c}{\cref{F4}(c) $\varphi_B = 0$} & \multicolumn{4}{c}{\cref{F4}(d) $\varphi_B=0.9$} \\
 $\varphi_L$ & $\varphi_R$ & $n_g$ & $\chi$ & $\varphi_L$ & $\varphi_R$ & $n_g$ & $\chi$ \\ \hline
 3.7806 & 3.6014 & 0.2598 & $-1$ & 3.5632 & 3.2152 & 0.2541 & $-1$ \\
 2.5026 & 2.6818 & 0.2598 & $+1$ & 1.9204 & 2.0659 & 0.4211 & $+1$ \\
 2.5026 & 2.6818  & 0.7403 & $+1$ & 1.9204 & 2.0659 & 0.5789 & $+1$ \\
 3.7806 & 3.6014 & 0.7403 & $-1$ & 3.5632 & 3.2152 & 0.7459 & $-1$ \\
\end{tabular}
\end{ruledtabular}
\end{table*}



\begin{figure}
\includegraphics[width=\columnwidth]{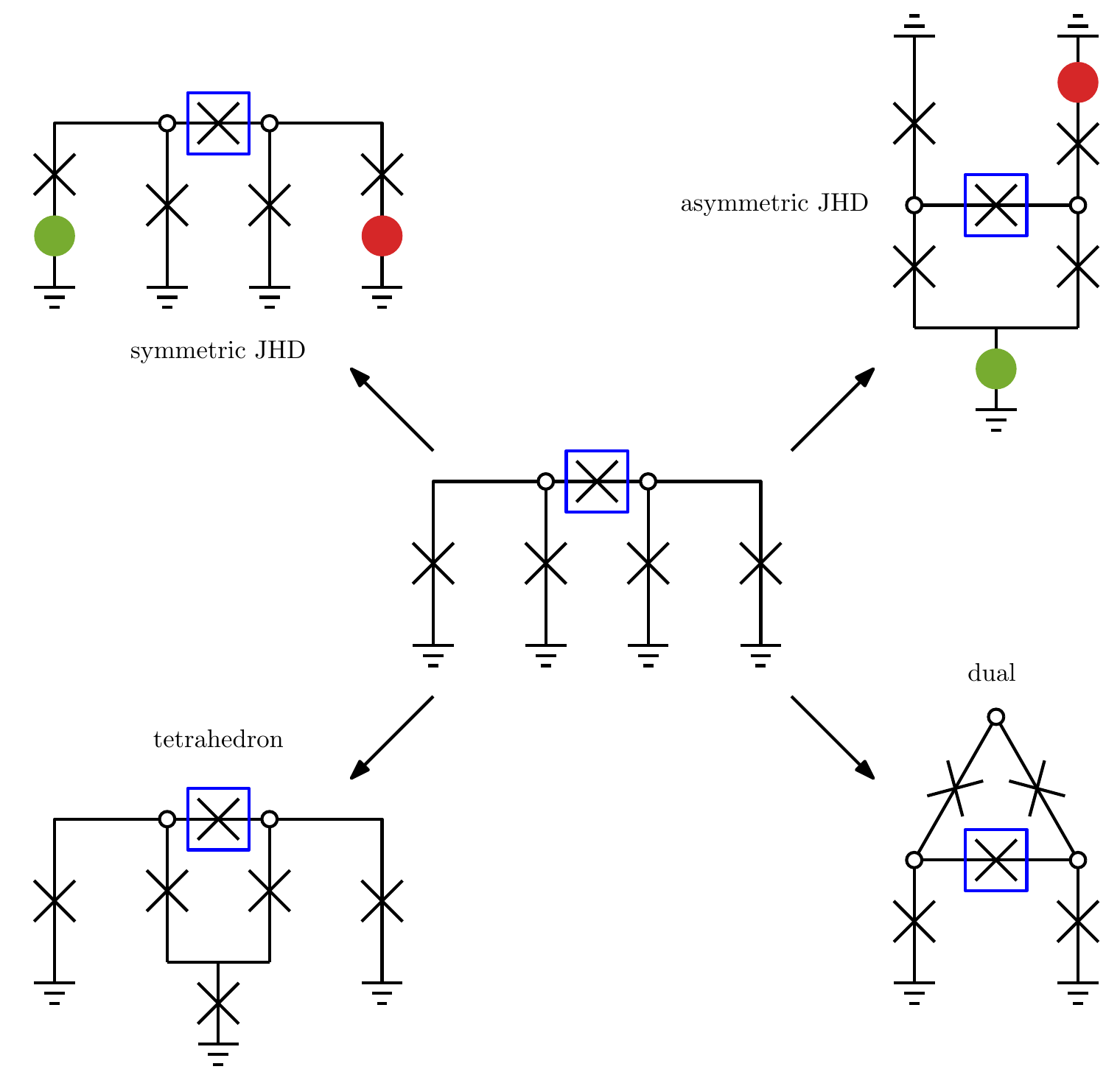}%
\caption{\label{FA2}\textbf{Correspondence between the symmetric and asymmetric Josephson quantized Hall conductance devices, the dual JHD circuit, and the tetrahedron circuit.}
  The elementary five junction JHD circuit shown in~\cref{F1} (center) can be mapped to the circuit of~\cref{F4} (upper left) by adding voltage sources (red and green circles).
  To obtain the circuit of~\cref{F3} the green source is inserted as shown in the upper right and the remaining circuit is folded upwards.
  The tetrahedron circuit (lower left) requires an additional junction.
  The JHD dual circuit (lower right) is constructed from the dual graph of the central circuit after connecting grounds.
  Whereas the JHD has two islands (unfilled circles) and three loops, the dual circuit has three islands and two loops.
  Although transconductance is quantized in the JHD and tetrahedron, we have not verified that this is true for the dual JHD circuit.}
\end{figure}


\begin{figure*}[h]
\includegraphics[width=0.8\textwidth]{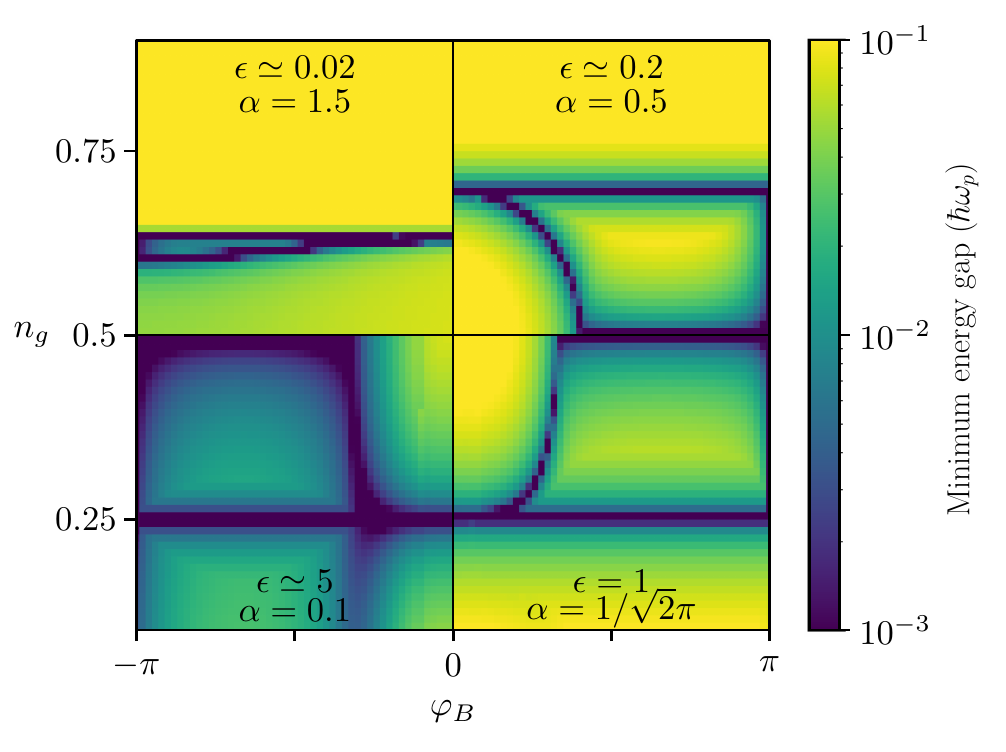}
\caption{\label{FA3}\textbf{Optimizing internal parameters of the Josephson quantized Hall conductance device.}
  The minimum energy gap diagram of~\cref{F5} is reproduced for identical junctions and with each quadrant corresponding to a different value of the normalized junction impedance $\alpha = Z_J/R_Q = 1/2\pi\cdot\sqrt{2E_C/E_J}$ or energy ratio $\epsilon = E_J/E_C$.
  For a given value of $\alpha$ the quadrants not shown are related by symmetry.
  Energy is plotted in units of the plasma frequency $\hbar\omega_p = \sqrt{2E_JE_C}$ which is constant so that the gap can be compared for different values of $\alpha$, and gap values beyond the color scale are capped.
  The upper left corresponds to the deep charging regime $E_C \gg E_J$ and the Chern number is zero everywhere except in the top left pocket.
  The degeneracies are located almost entirely on the horizontal line at $n_g \approx 5/8$ (and $n_g \approx 3/8$ by symmetry) as expected for a Cooper pair pump in which the exterior capacitances are doubled.
  As the Josephson energy is increased clockwise, the topologically non-trival region grows out from the corner pockets where these horizontal lines reach $\varphi_B = \pm\pi$.
  An additional horizontal degeneracy line appears at $n_g=1/2$ and the ones at $n_g = 2/3$ ($n_g=1/3$) move out towards $n_g = 3/4$ ($n_g = 1/4$).
  From the color scale, the minimum energy gap is maximized near $\alpha = 0.5$ (upper right).
  By designing the junction area $S \propto 1/\alpha$ for the optimal impedance, the energy gap in the topologically non-trivial region is maximized, reducing error in transconductance quantization due to Landau-Zener transitions.
  The bottom right quadrant ($\epsilon = 1$) can be compared to~\cref{F5}, where the junctions are not identical but the average $E_{Ji}/E_{Ci}$ is approximately one. 
  As junction uniformity is reduced, the corners of the non-trivial region are pulled back.
}
\end{figure*}

\clearpage 
\bibliography{main}

\end{document}